\documentclass[journal,twoside,web]{ieeecolor}
\usepackage{tmi2}
\usepackage{cite}
\usepackage{amsmath,amssymb,amsfonts}
\usepackage[ruled,vlined]{algorithm2e}
\usepackage{graphicx}
\usepackage{textcomp}
\def\BibTeX{{\rm B\kern-.05em{\sc i\kern-.025em b}\kern-.08em
    T\kern-.1667em\lower.7ex\hbox{E}\kern-.125emX}}
\usepackage{tikz}
\usetikzlibrary{matrix,calc}

\definecolor{brightcerulean}{rgb}{0.11, 0.62, 0.74}

\usepackage{comment}
\usepackage{amsmath}
\usepackage{amssymb}
\usepackage{latexsym}
\usepackage{framed,multirow}
\usepackage{url}
\usepackage{xcolor}
\usepackage{lipsum}
\usepackage{hyperref}
\usepackage{graphics}
\usepackage{gensymb}
\SetKwInput{KwInput}{Input}                % Set the Input
\SetKwInput{KwOutput}{Output}              % set the Output
\hypersetup{
    colorlinks=true,
    linkcolor=blue,
    filecolor=magenta,      
    urlcolor=cyan,
    pdftitle={DreaMR}
    }

\definecolor{newcolor}{rgb}{.8,.349,.1}
\def\arraystretch{1.4} % Vertical Padding
\usepackage[font=small,justification=justified,belowskip=2pt,aboveskip=2pt]{caption}
\usepackage[math]{cellspace}
\usepackage{multirow}
\usepackage{nicematrix}
\usepackage{makecell}

\begin{document}
\title{DreaMR: Diffusion-driven Counterfactual Explanation for Functional MRI}
\author{Hasan A. Bedel and Tolga \c{C}ukur$^*$ %\IEEEmembership{Senior Member} 
\vspace{-1.4cm}
\\
\thanks{This study was supported in part by a TUBITAK BIDEB scholarship awarded to H. A. Bedel, by TUBA GEBIP 2015 and BAGEP 2017 fellowships, and by a TUBITAK 1001 Grant 121N029 awarded to T. \c{C}ukur  (Corresponding author: Tolga \c{C}ukur).}
\thanks{H. A. Bedel and T. \c{C}ukur are with the Department of Electrical and Electronics Engineering, and the National Magnetic Resonance Research Center (UMRAM), Bilkent University, Ankara, Turkey, 06800. (e-mails: abedel@ee.bilkent.edu.tr, cukur@ee.bilkent.edu.tr).  T. \c{C}ukur is also with the Neuroscience Graduate Program Bilkent University, Ankara, Turkey, 06800. }
}

\maketitle
\begin{abstract}
Deep learning analyses have offered sensitivity leaps in detection of cognitive states from functional MRI (fMRI) measurements across the brain. Yet, as deep models perform hierarchical nonlinear transformations on their input, interpreting the association between brain responses and cognitive states is challenging. Among common explanation approaches for deep fMRI classifiers, attribution methods show poor specificity and perturbation methods show limited plausibility. While counterfactual generation promises to address these limitations, previous methods use variational or adversarial priors that yield suboptimal sample fidelity. Here, we introduce the first diffusion-driven counterfactual method, DreaMR, to enable fMRI interpretation with high specificity, plausibility and fidelity. DreaMR performs diffusion-based resampling of an input fMRI sample to alter the decision of a downstream classifier, and then computes the minimal difference between the original and counterfactual samples for explanation. Unlike conventional diffusion methods, DreaMR leverages a novel fractional multi-phase-distilled diffusion prior to improve sampling efficiency without compromising fidelity, and it employs a transformer architecture to account for long-range spatiotemporal context in fMRI scans. Comprehensive experiments on neuroimaging datasets demonstrate the superior specificity, fidelity and efficiency of DreaMR in sample generation over state-of-the-art counterfactual methods for fMRI interpretation.
\vspace{-0.1cm}
\end{abstract}

% keywords can be removed
\begin{IEEEkeywords} counterfactual, explanation, interpretation, generative, diffusion, functional MRI \vspace{-0.4cm}
\end{IEEEkeywords}

\bstctlcite{IEEEexample:BSTcontrol}

\section{Introduction}
Functional magnetic resonance imaging (fMRI) enables non-invasive cognitive assessment by recording blood-oxygen-level-dependent (BOLD) responses across the brain consequent to neural activity \cite{singleton2009functional}. Spatiotemporal BOLD responses measured under various cognitive tasks can be analyzed to infer associations between brain regions and individual cognitive states. A traditional framework in neuroscience performs inference via linear classifiers that are trained to predict cognitive state given responses \cite{nishimoto2011}. The classifier weight for a brain region is then taken to reflect the importance of that region in mediating the respective cognitive state. Although this traditional approach offers ease of interpretation, linear classifiers typically suffer from poor sensitivity \cite{pereira2009machine}. In recent years, deep-learning classifiers have gained prominence as they show substantially higher sensitivity to fine-grained patterns in fMRI data \cite{jang2017task,huang2017modeling,kawahara2017brainnetcnn,parisot2018disease,kam2019deep,li2020deep,kim2021learning,wang2021graph}. Despite this important benefit, hierarchical layers of nonlinear transformation in deep models obscure precise associations between brain responses and cognitive states, introducing an interpretation challenge and creating a barrier to methodological trust \cite{vandervelden2022XAI}. As such, there is a dire need for explanation methods that highlight the critically important set of input features (i.e., brain responses) for deep fMRI models to help interpret their decisions. 

An emerging explanation framework is based on generation of counterfactual samples \cite{vandervelden2022XAI}. Counterfactual explanation aims to identify a minimal, plausible set of changes in the features of an input fMRI sample that is sufficient to alter the decision of a downstream analysis model \cite{matsui2022counterfactual}. A generative prior is commonly used to produce a counterfactual sample proximal to the original sample in the domain of the data distribution \cite{rodriguez2021beyond}. The difference between the two samples is then inspected to interpret associations between brain regions and cognitive state \cite{matsui2022counterfactual}. Counterfactual methods have been reported to attain superior feature specificity against attribution methods, which derive gradient or activation heatmaps that can be broadly distributed across input features \cite{kim2020understanding,tomaz2021visual}. They can also produce more plausible interpretations against perturbation methods, which perform local degradations on input features that can disrupt global coherence \cite{kazemi2018deep, devereux2018integrated}. Nevertheless, the performance of counterfactual methods depend critically on the fidelity of samples synthesized by the underlying generative prior. 

Previous studies on counterfactual explanation have commonly proposed variational autoencoder (VAE) or generative adversarial network (GAN) priors trained to capture the distribution of fMRI data \cite{matsui2022counterfactual,rodriguez2021beyond,cohen2021gifsplanation}. These priors synthesize counterfactuals with high efficiency, albeit VAEs often suffer from relatively low sample quality due to loss of detailed features, and GANs suffer from training instabilities that can hamper sample quality or diversity \cite{korkmaz2022unsupervised}. A promising surrogate for sample generation is diffusion priors that offer high sample fidelity via a many-step sampling process \cite{ho2020denoising,song2020score}. Few recent imaging studies have considered diffusion-based counterfactual generation to detect anomalous lesions in anatomical MRI and X-ray scans \cite{sanchez2022healthy,wolleb2022mad,pinaya2022mad}. Yet, to our knowledge, diffusion priors have not been explored for counterfactual explanation in multi-variate fMRI analysis. This remains a challenge given the computational burden of sampling with conventional diffusion priors such as DDPM or DDIM that require hundreds of sampling steps \cite{song2020denoising}, coupled with the high intrinsic dimensionality of fMRI data.

\begin{figure*}[th]
\vspace{-0.1cm}
  \centering
\includegraphics[width=0.85\linewidth]{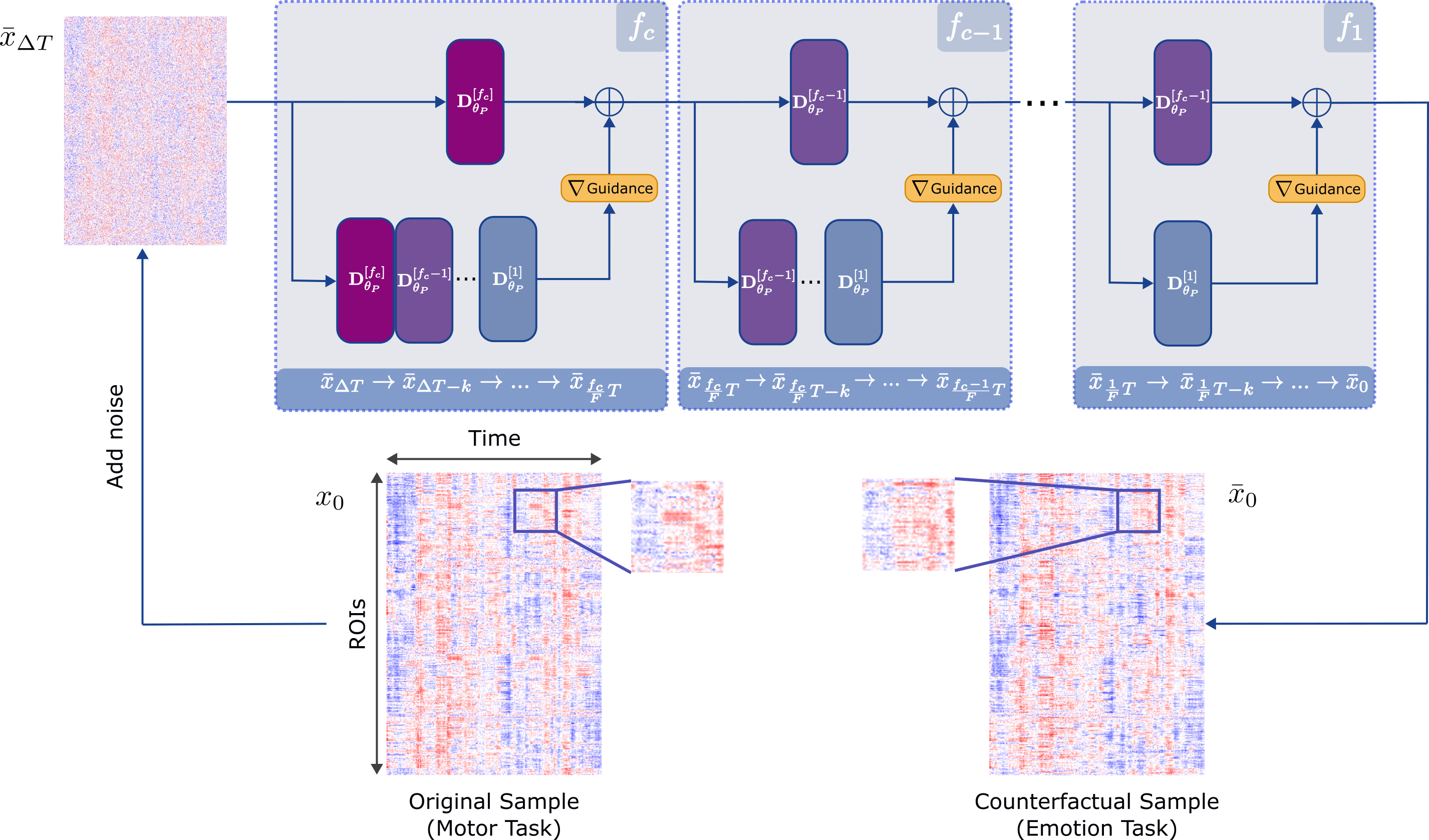}  
  \caption{DreaMR is a diffusion-driven counterfactual explanation method for deep fMRI models. Given a classifier that maps an input sample $x_0$$\sim$$p(x)$ onto a cognitive state $y_0$, DreaMR first samples a noise-added version $\Bar{x}_{\Delta T}$ via forward diffusion and then generates $\Bar{x}_0$$\sim$$p(x)$ with minimal alterations from $x_0$ via reverse diffusion. Generation is guided with the conditional score of the classifier to flip the decision to $\Bar{y}_0 \neq y_0$. For effective and efficient control over generated samples, a novel FMD prior is employed that splits the diffusion process into $F$ uniform fractions with dedicated networks $\mathbf{D}_{\theta_P}^{[f]}$. Generation starts at fraction $f_c=\lceil \Delta T (F/T) \rceil$, and  classifier guidance computed from a denoised estimate of the counterfactual sample is injected at the end of each fraction (orange boxes).}
\label{fig:DreaMR}
\end{figure*}

Here, we propose a novel diffusion-driven counterfactual explanation method, DreaMR, to interpret downstream fMRI classifiers with improved fidelity and efficiency. The proposed method trains a class-agnostic diffusion prior for fMRI data, and the trained prior generates a counterfactual sample with guidance from a downstream classifier to alter its decision (Fig. \ref{fig:DreaMR}). To improve efficiency without compromising sample quality, DreaMR leverages a novel fractional multi-phase-distilled diffusion (FMD) prior that splits the diffusion process into consecutive fractions and performs multi-phase distillation in each fraction (Fig. \ref{fig:FMD}). Unlike regular diffusion priors based on UNet backbones, DreaMR leverages an efficient transformer architecture of linear complexity to capture long-range spatiotemporal context in fMRI scans. During counterfactual generation, classifier-guidance is injected separately into each diffusion fraction to tailor synthesis of intermediate samples in proximity of the original sample. The difference between the original and the final counterfactual samples reflects the contribution of brain regions to the model decision. 

We report comprehensive demonstrations to explain deep classifiers for gender detection on resting-state fMRI scans, and to explain deep classifiers for cognitive task detection in task-based fMRI scans. We find that DreaMR achieves superior fidelity against competing explanation methods, significantly outperforming conventional diffusion priors in computational efficiency. Code for DreaMR will be available at {\small \url{https://github.com/icon-lab/DreaMR}}.

\vspace{0.15cm}
\subsubsection*{\textbf{Contributions}}
\begin{itemize}
    \item To our knowledge, we introduce the first diffusion-driven counterfactual explanation method in the literature for multi-variate fMRI analysis.
    \item DreaMR generates counterfactual samples with a novel fractional multi-phase-distilled diffusion prior to boost sampling efficiency without compromising quality.
    \item Unlike conventional diffusion methods, DreaMR leverages a transformer architecture of linear complexity to capture long-range spatiotemporal context in fMRI scans. 
\end{itemize}

\section{Related Work}
\subsection{Explanation of fMRI Models}
The complex nature of multivariate fMRI data presents an opportunity for deep learning with its superior sensitivity over traditional approaches \cite{jang2017task,huang2017modeling,kawahara2017brainnetcnn,parisot2018disease,kam2019deep,li2020deep}. At the same time, the hierarchy of nonlinear transformations in deep models makes it challenging to interrogate how they render decisions, limiting interpretability. Two prominent frameworks have been proposed in the fMRI literature to address this challenge. A main framework is intrinsic interpretation where analyses are conducted using specialized models, such as linearized or graph classifiers, with restricted designs to permit an inherent degree of explanation \cite{ktena2018metric,wang2021graph,sivgin2023plugin}. Model-specific explanation is then performed by inspecting internal parameters, but this framework is limited to relatively simpler models that often underperform against state-of-the-art deep models \cite{malkiel2021pre,yu2022disentangling,zhang2022diffusion,bedel2023bolt}. 

A more flexible framework is post-hoc explanation where fMRI analyses are conducted using a preferred model without restrictions, and interpretation is achieved by observing the influence of model inputs on the output. Among common techniques, attribution methods derive heatmaps across input features to estimate their salience, taken as gradient of the target class with respect to input features \cite{li2020multi,riaz2020deepfmri}, activations for the target class \cite{ellis2022approach,gotsopoulos2018reproducibility, vu2020fmri, arslan2018graph}, or a weighted combination of gradients and activations \cite{kim2020understanding,lin2022sspnet,tomaz2021visual}. Attribution methods often require architecture-based modifications that can limit practicality, and they tend to produce broadly distributed heatmaps that degrade interpretation specificity. Meanwhile, perturbation methods introduce patch-level degradation on input features through operations such as occlusion and inpainting \cite{huang2017modeling, wang2021graph, devereux2018integrated, kazemi2018deep,li2020hypernetwork, hu2020interpretable, dos2023assessing}. While such degradations can improve local specificity, perturbation-methods often produce globally-incoherent samples that hamper the plausibility of interpretations. 

Counterfactual explanation methods are a powerful alternative that produce specific results by identifying a minimally-sufficient set of changes to input samples to flip the model decision, and ensure plausible results by resampling input samples via a generative prior on fMRI data \cite{matsui2022counterfactual}. That said, VAE and GAN priors in previously proposed counterfactual methods for fMRI can suffer from loss of structural details or low fidelity in counterfactual samples \cite{vandervelden2022XAI}. Recent machine learning studies advocate diffusion priors as a promising surrogate for reliable sample generation, albeit conventional diffusion priors are known to suffer from poor sampling efficiency \cite{ho2020denoising,song2020score}. This impedes adoption of diffusion priors in counterfactual generation that requires iterated resampling of each high-dimensional fMRI data sample until the respective model decision is flipped. To address these open issues, here we introduce DreaMR, the first diffusion-driven explanation method for multivariate fMRI analysis to our knowledge.

\subsection{Counterfactual Generation}
To improve reliability in counterfactual generation, recent computer vision studies have considered diffusion priors to synthesize natural images from desired object classes. \cite{augustin2022diffusion,sanchez2022diffusion,jeanneret2022diffusion} propose conventional diffusion priors based on a common UNet architecture. Interleaved sampling is used to trade-off sample quality in return for efficiency \cite{sanchez2022diffusion,jeanneret2022diffusion}. Unlike conventional diffusion priors that typically require hundreds of diffusion steps even with interleaved sampling, DreaMR improves practicality by employing a novel diffusion prior (FMD) that simultaneously maintains high sampling efficiency and quality. To do this, FMD introduces a multi-phase distillation procedure on consecutive fractions of the diffusion process, where each fraction uses a dedicated denoising network. DreaMR uniquely implements denoising networks via a transformer architecture to improve sensitivity to long-range context in fMRI scans that last several minutes \cite{liegeois2019resting}. 

Few recent imaging studies have also considered diffusion-based counterfactual generation to map lesions in anatomical scans by synthesizing pseudo-healthy medical images \cite{sanchez2022healthy,wolleb2022mad,pinaya2022mad}. Commonly, these studies propose conventional diffusion priors based on UNet. Instead, DreaMR leverages the novel FMD prior for higher efficiency and a transformer architecture to capture long-range context. Note that \cite{sanchez2022healthy,wolleb2022mad} train class-conditional priors on normals and patients, and \cite{pinaya2022mad} trains a class-specific prior on normals. When adopted for counterfactual explanation, such class-informed priors must use matching class definitions to the classifier, so they require retraining for each classification task. In contrast, DreaMR employs a class-agnostic diffusion prior that can be utilized to explain models for different tasks without retraining.

\section{Theory}
\subsection{Conventional Diffusion Priors}
Diffusion priors use a gradual process to transform between a data sample $x_0$ and a random noise sample $x_T$ in T steps. In the forward direction, Gaussian noise is added to obtain a noisier sample $x_t$ at step $t$, with forward transition probability:
\begin{equation}
q\left( x_{t}|x_{t-1} \right)=\mathcal{N}\left( x_{t}; (\alpha_{t}/\alpha_{t-1})x_{t-1},\sigma^2_{t|t-1}\mathrm{I} \right)
\end{equation}
where $\mathcal{N}$ is a Gaussian distribution, $\mathrm{I}$ is the identity covariance matrix, and $\alpha$, $\sigma^2$ are scaling and noise variance parameters where $\alpha_t^2 + \sigma_t^2 = 1$ \cite{ho2020denoising}. In the reverse direction, a network parametrization $\mathbf{D_{\theta}}$ is used to restore original data from noisy samples, i.e., $\hat{x}_0=\mathbf{D_{\theta}}(x_t)$. The prior can be trained by minimizing a variational bound on likelihood \cite{song2020score}: 
\begin{eqnarray}
    \mathbb{E}_{t \sim U[1,T], x_0 \sim p(x), x_t \sim q(x_t|x_0)} \left[ \omega \left(\lambda_t \right) 
    \| \mathbf{D_{\theta}}(x_t)-x_0 \|_2^2
    \right]
\end{eqnarray}
where $\mathbb{E}$ is expectation, $U$ is a uniform distribution, $q(x_t|x_0) = \mathcal{N}\left(x_t; \alpha_tx_0, \sigma_t^2\mathrm{I} \right)$, $\omega\left(\lambda_t\right)$ is a weighting function with $\lambda_t = log(\alpha_t^2 / \sigma_t^2)$ denoting signal-to-noise ratio. 

Once trained, the diffusion prior can generate a synthetic data sample by progressively denoising a random noise sample across $T$ steps. The reverse transition probability for the denoising steps can be expressed as \cite{ho2020denoising}:
\begin{eqnarray}
&&q\left( x_{t-1} | x_t, \hat{x}_0 \right) = \notag \\ 
&&\mathcal{N} \left( \alpha_{t-1}\hat{x}_0 + \sqrt{1- \alpha_{t-1}^2 - \beta_t^2} \; . \frac{x_t - \alpha_t \hat{x}_0}{\sigma_t}, \; \beta_t^2 \mathrm{I} \right) 
\end{eqnarray}
where $\beta_t = \frac{\sigma_{t-1}}{\sigma_t \; \alpha_{t-1}} \sqrt{(\alpha_{t-1}^2 - \alpha_t^2)}$ controls the stochasticity of generated samples. The original diffusion prior requires $T$$\approx$1000 forward passes through the network for generation. To lower sampling time, a common solution is interleaved sampling with step size $k$, $x_{t-k}$$\sim$$q(x_{t-k}|x_t,\hat{x}_0)$, while $\beta_t = 0$ for deterministic generation (i.e., DDIM) \cite{song2020denoising}.
Typically, interleaved sampling still requires few hundred steps for generation and it can suffer from reduced sample quality \cite{sanchez2022diffusion}.

\subsection{DreaMR}

\begin{figure*}[th]
\vspace{-0.1cm}
  \begin{minipage}{0.33\textwidth}
     \caption{DreaMR leverages a novel FMD prior for efficient sample generation without compromising sample quality, and to enable effective intermittent control over the generation process. The FMD prior splits the overall diffusion process into $F$ fractions, where the $f$th fraction covers $T/F$ consecutive steps from $t_{s}(f)=T(f)/F$ to $t_{e}(f)=T(f-1)/F+1$. To improve sample fidelity, a dedicated denoising network $\mathbf{D}_{\theta_P}^{[f]}$ is employed in each fraction. To enhance sensitivity to long-range temporal context in fMRI scans, the network is built on an efficient transformer architecture that uses fused window attention mechanisms \cite{bedel2023bolt}. To improve efficiency without compromising sample quality, multi-phase distillation is performed over $P$ phases in each fraction. In the $p$th distillation phase, the diffusion step size is doubled to shorten sampling time by a factor of 2. At the end of multi-phase distillation, the number of sampling steps that must be executed for a given fraction is reduced from $T/F$ to $T/(2^P F)$. }
     \label{fig:FMD}
  \end{minipage}
  \begin{minipage}{0.67\textwidth}
     \centering
     \includegraphics[width=0.95\columnwidth]{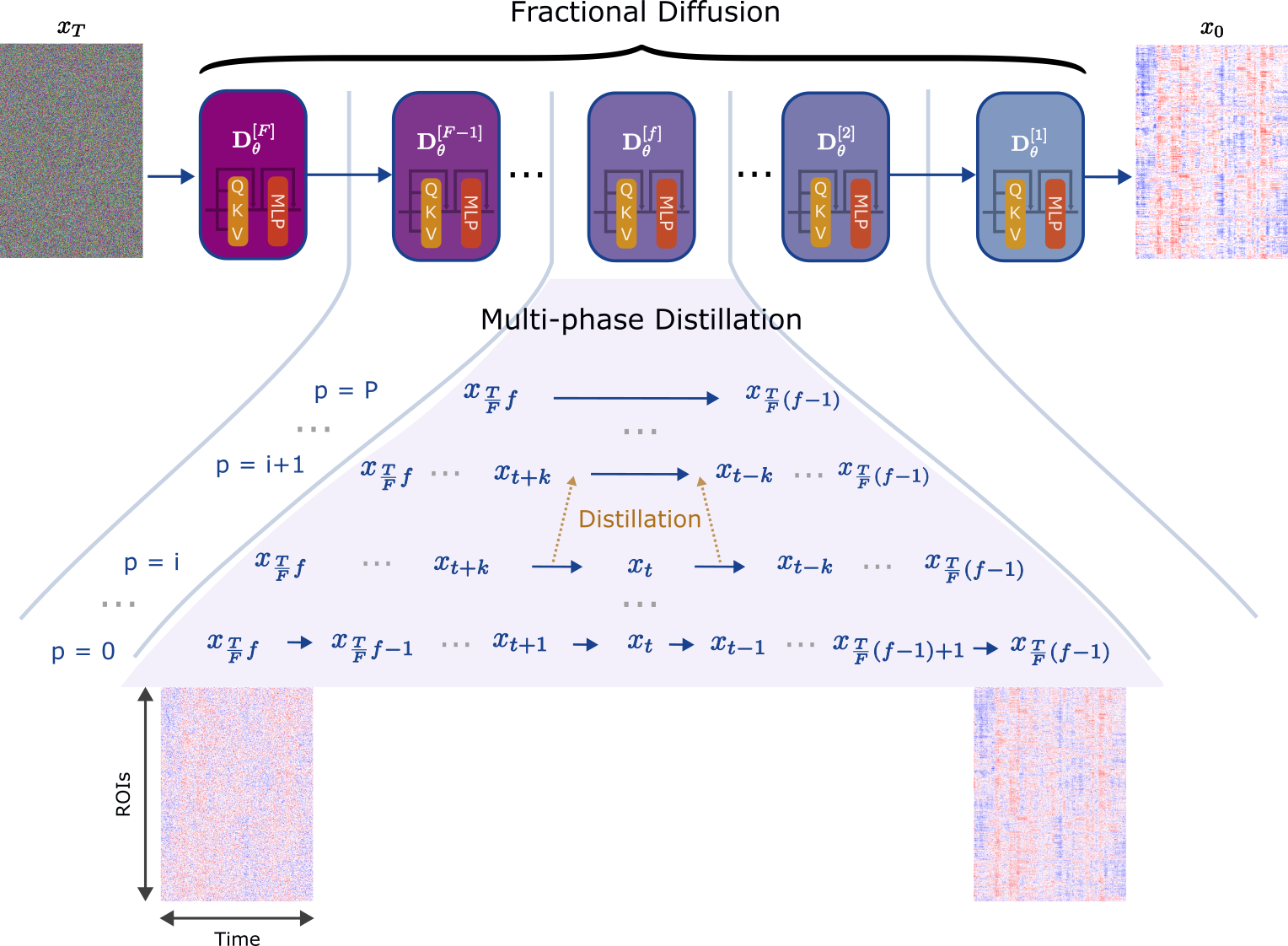}  
  \end{minipage}
  \hfill
\end{figure*}

DreaMR is a novel explanation method for deep fMRI models based on counterfactual generation. Assume that a downstream classifier $c(x)$$=$$y$ maps an input fMRI sample $x_0$$\sim$$p(x)$ onto a class label $y_0\in Y$ for cognitive state, according to posterior probability $p_c(y|x)$. Counterfactual generation aims to obtain plausible samples $\Bar{x}_0$$\sim$$p(x)$ with minimal alterations from $x_0$, such that the classifier decision is flipped $c(\Bar{x}_0) = \Bar{y}_0$ where $\Bar{y}_0 \neq y_0, \Bar{y}_0 \in Y$. Afterwards, the differences between the original and counterfactual samples ($x_0$–$\Bar{x}_0$) can be inspected to infer the input features that are critical in distinguishing between cognitive states $y_0$ and $\Bar{y}_0$. 

Diffusion priors hold great promise for counterfactual generation since proximal variants of $x_0$ can be obtained via cyclic Langevin sampling \cite{augustin2022diffusion}. To do this, a noise-added version $\Bar{x}_{\Delta T}$ is first sampled via the forward process:
\begin{equation}
    \label{eq:counter_vanilla_forward}
    q\left( \Bar{x}_{\Delta T}|x_{0} \right)=\mathcal{N}\left( \Bar{x}_{\Delta T}; (\alpha_{\Delta T}/\alpha_{0})x_{0},\sigma^2_{\Delta T|0}\mathrm{I} \right)
\end{equation}
Starting reverse diffusion at step $\Delta T$, a counterfactual can then be generated via guidance from the classifier \cite{sanchez2022diffusion}:
\begin{equation}
    \label{eq:counter_vanilla_backward}
    q(\Bar{x}_{t-1}|\Bar{x}_t, \hat{\Bar{x}}_0) = \mathcal{N} \left( \hat{\Bar{x}}_0 + s \, \beta_t^2 \, \nabla_{\Bar{x}_t} log \, p_c(\Bar{y}_0|\Bar{x}_t) , \beta_t^2 \right)
\end{equation}
where $\nabla_{\Bar{x}_t}$ is gradient with respect to $\Bar{x}_t$, $s$ is a scaling parameter. While previous studies on counterfactual generation have proposed interleaved sampling for acceleration, generation can still last few hundred steps with conventional diffusion priors \cite{song2020denoising}. Furthermore, $\nabla_{\Bar{x}_t} log \, p_c(\Bar{y}|\Bar{x}_t)$ is unknown as the classifier is trained on original samples without added noise. 

To improve practicality, DreaMR leverages a novel FMD prior for efficient generation in few steps without compromising sample quality (Fig. \ref{fig:DreaMR}). Classifier guidance is injected at the end of each fraction to effectively guide generation. To avoid the need for classifier retraining, DreaMR computes classifier gradients given denoised sample estimates as a surrogate for gradients on noisy samples in each fraction. The working principles are detailed in the remainder of the section.

\subsubsection{FMD prior}
With a conventional diffusion prior, providing classifier guidance at the final time step can suffer from suboptimal control over generated samples, whereas providing guidance at each diffusion step involves calculation of classifier gradients that can be computationally prohibitive \cite{sanchez2022healthy}. To enable effective control without alleviating computational burden, we propose a novel FMD prior that splits the diffusion process into $F$ uniform fractions, such that classifier guidance can be injected intermittently at the end of each fraction (Fig. \ref{fig:FMD}). The $f$th fraction covers $T/F$ consecutive steps from $t_{s}(f)=T(f)/F$ to $t_{e}(f)=T(f-1)/F+1$, and employs a dedicated denoising network $\hat{x}_{0} = \mathbf{D}_{\theta}^{[f]}(x_t)$. The FMD prior is trained via the following objective:
\begin{eqnarray}
   \sum_{f=1}^{F} \mathbb{E}_{t \sim U[t_s(f),t_e(f)]} \left[ 
    \| \mathbf{D}_{\theta}^{[f]}(x_t)-x_0 \|_2^2
    \right]
\end{eqnarray}
where $\omega$ and expectation over $x_{0}$, $x_{t}$ are omitted for brevity. 

To further improve efficiency, FMD leverages multi-phase distillation for accelerated sampling as inspired by a recent computer vision study \cite{salimans2022progressive}. Yet, \cite{salimans2022progressive} uses a common distillation across the entire diffusion process, which can compromise fidelity due to varying noise levels, feature details across diffusion steps \cite{balaji2022ediffi}. Instead, FMD leverages fraction-specific distillations to maintain high sample quality (Fig. \ref{fig:FMD}). To minimize information loss during knowledge transfer, the original teacher network $\mathbf{D}_{\theta_0}^{[f]}(x_t)$ is gradually distilled onto a student network $\mathbf{D}_{\theta_P}^{[f]}$ over $P$ phases. In the $p$th phase, $\mathbf{D}_{\theta_{p-1}}^{[f]}(x_t)$ is the teacher, $\mathbf{D}_{\theta_{p}}^{[f]}(x_t)$ is the student, and the diffusion step size is increased by a factor of 2 as follows:
\begin{eqnarray}
    \mathbb{E}_{t \sim U(\{t_s(f):T/(k_d F):t_e(f)\})}\left[ \| \mathbf{D}^{[f]}_{\theta_p}(x_t)-\Tilde{x}_{0} \|_2^2
    \right] \label{eq:distill_obj}
\end{eqnarray}
where $k_d = 2^{p}$ is diffusion step size of the student. Adopting interleaved sampling, the reference sample $\Tilde{x}_{0}$ is derived via sampling based on the teacher $\mathbf{D}_{\theta_{p-1}}^{[f]}(x_t)$:
\begin{equation}
    \Tilde{x}_{0} = \frac{\Tilde{x}_{t-2k_{o}} - (\sigma_{t-2k_{o}} / \sigma_t)x_t}{\alpha_{t-2k_{o}} - (\sigma_{t-2k_{o}} / \sigma_t)\alpha_t} 
\end{equation}
where $k_{o} = 2^{(p-1)}$ is the diffusion step size of the teacher. At the end of the distillation process, the number of steps for a given fraction is reduced from $T/F$ to $T/(2^P F)$.

\subsubsection{Counterfactual generation}
A counterfactual sample that flips the classifier decision can be drawn if the joint score function $\nabla_{x_t} log \, p(x_t, y)$ for noisy samples and predicted class labels is known. Since $p(x_t, y) = p(x_t)p_c(y|x_t)$ based on Bayes' rule, the joint score is given as: 
\begin{align}
    \nabla_{x_t} log \, p(x_t, y) &= \nabla_{x_t} log \, p(x_t) + \nabla_{x_t} log \, p_c(y|x_t)
\end{align}
The first term denotes the marginal score function for noisy samples, which can be derived as follows \cite{ho2020denoising}:
\begin{align}    
    &p(x_t) = \frac{1}{\sigma_t \sqrt{2\pi}} exp \left( -\frac{1}{2} 
    \left( \frac{x_t - \alpha_t \hat{x}_0}{\sigma_t} \right)^2 \right) \\
    &\nabla_{x_t} log \, p(x_t) = -\frac{x_t - \alpha_t \hat{x}_0 }{\sigma_t^2}
\end{align}
Meanwhile, the second term denotes conditional score function for predicted labels given noisy samples. Note that the originally trained classifier does not capture $p_c(y|x_t)$, but instead $p_c(y|x_0)$. The classifier could be retrained on a set of noisy samples generated by the diffusion prior. However, this brings in additional computational burden, and learning of $p_c(y|x_t)$ on samples with heavy noise might be difficult. 

To avoid limitations related to classifier retraining, here we derive a surrogate based on $p_c(y|x_0)$ to compute the conditional score function. Following the chain rule:
\begin{align}
         \nabla_{x_t} log \, p_c(y|x_t) &= \nabla_{\Tilde{x}_0} log \, p_c(y|x_t) \cdot \nabla_{x_t} \Tilde{x}_0
\end{align}
where $\Tilde{x}_0$ is obtained via sampling across the diffusion process. Based on forward diffusion, $x_t$ and $x_0$ are related as:
\begin{equation}
\label{eq:forward_diff}
x_{t}=\alpha_{t}x_0 +\sigma_{t}\epsilon,\mbox{ s.t. }\epsilon\sim \mathcal{N}\left( 0,\mathrm{I} \right)
\end{equation}
Assuming that $\Tilde{x}_0 \simeq x_0$, $\nabla_{x_t} \Tilde{x}_0 = 1/\alpha_t$. In conventional diffusion priors with stochastic sampling (i.e., $\beta_t \neq 0$), $x_t$ and $\Tilde{x}_0$ can be related through a one-to-many mapping. Yet, since DreaMR adopts deterministic sampling with $\beta_t = 0$, a unique $\Tilde{x}_0$ is obtained given $x_t$. As such, $p_c(y|x_t) = p_c(y|\Tilde{x}_0)$, and the conditional score can be expressed as:
\begin{align}
\label{eq:guidance}
         \nabla_{x_t} log \, p_c(y|x_t) &= \frac{1}{\alpha_t} \nabla_{\Tilde{x}_0} log \, p_c(y|\Tilde{x}_0) 
\end{align}
Since $p_c(y|\Tilde{x}_0)$$\,\simeq\,$$p_c(y|x_0)$ for a reasonably well-trained diffusion prior, the posterior distribution of the originally trained classifier can be used to provide guidance without retraining.  

DreaMR generates counterfactual samples by injecting intermittent classifier guidance according to Eq. \ref{eq:guidance} at the end of each fraction (see Fig. \ref{fig:DreaMR} and Alg. \ref{algorithm:DreaMR_algo}). Given an original fMRI sample $x_0$, a noisy sample $\bar{x}_{\Delta T}$ is first obtained via forward diffusion as in Eq. \ref{eq:forward_diff}. Counterfactual samples are then generated via reverse sampling across the fractions. For the $f$th fraction, the sample $\bar{x}_{t_s(f)}$ at the beginning of the fraction is subjected to interleaved sampling until the last step: 
\begin{equation}
  \bar{x}_{t-k} = \alpha_{t-k} \hat{\bar{x}}_0 + \sigma_{t-k} \frac{\bar{x}_t - \alpha_t \hat{\bar{x}}_0}{\sigma_t}  
\end{equation}
where $k$ is step size, and $\hat{\bar{x}}_0 = \mathbf{D}_{\theta_P}^{[f]}(\bar{x}_t)$. The sampling at the last step is injected with classifier guidance to elicit label $\bar{y}$:
\begin{eqnarray}
\label{eq:cgen}
    \bar{x}_{t-k} &=& \alpha_{t-k} \hat{\bar{x}}_0 + \sigma_{t-k} \frac{\bar{x}_t - \alpha_t \hat{\bar{x}}_0}{\sigma_t} + \nonumber \\
    && s \frac{\sigma_t^2}{\alpha_t^2} (\alpha_{t-k} - \alpha_t \frac{\sigma_{t-k}}{\sigma_t}) \nabla_{\Tilde{\bar{x}}_0} log \, p_c(\bar{y}|\Tilde{\bar{x}}_0)
\end{eqnarray}
where $t = t_e(f)+k$, and $s$ is a scaling constant to control the strength of guidance. In Eq. \ref{eq:cgen}, the conditional score from the classifier is computed using $\Tilde{\bar{x}}_0$, which is Langevin sampled across the diffusion process, instead of $\hat{\bar{x}}_0$, as this was observed to improve the quality of guidance.

\begin{algorithm}[t]
\caption{Counterfactual generation with DreaMR} 
\label{algorithm:DreaMR_algo}
%\small
\SetAlgoLined
\KwIn{\\ $x_0 \sim p(x)$: Original fMRI sample \\ 
$p_c(y|x)$: Posterior probability of the classifier \\
$y_0 \in Y$: Classifier-predicted label for $x_0$ \\ 
$\Bar{y}_0$: Target label for counterfactual generation \\
$\Delta T$: Initial diffusion step for counterfactual generation \\
$\{ \mathbf{D}_{\theta_P}^{[1]}, ..., \mathbf{D}_{\theta_P}^{[F]}  \}$: Distilled networks across fractions \\
$k = 2^P$: Step size after P distillation phases
}
\KwOut{
   \\ $\Bar{x}_0$: Counterfactual sample 
}
$\bar{x}_{\Delta T} \leftarrow \alpha_{\Delta T}x_0 + \sigma_{\Delta T}\epsilon$ \hspace{0.1em} $\triangleright$ generate noisy sample \\
\For{$f_o$ in range($F$, 0, -1)}{
     {$\triangleright$ sample across fractions to find $\Tilde{\bar{x}}_{0}$} \\
    \For{$t_i$ in range($t_o$, $-1$, $-k$)}{ 
        $f_i \leftarrow ceil(t_i F/T)$ \\ 
        $\hat{\bar{x}}_{0} \leftarrow \mathbf{D}_{\theta_P}^{[f_i]}(\bar{x}_{t_i})$ \\
        $\bar{x}_{t_i-k} \leftarrow \alpha_{t_i-k} \hat{\bar{x}}_{0}   + \sigma_{t_i-k} \frac{\bar{x}_{t_i} - \alpha_{t_i} \hat{\bar{x}}_{0} }{\sigma_{t_i}}$
    }
    $\Tilde{\bar{x}}_{0} \leftarrow {\bar{x}}_{0} $; $ G \leftarrow \nabla_{\Tilde{\bar{x}}_0} log \, p_c(\Bar{y} \,|\, \Tilde{\bar{x}}_0)$ $\triangleright$ gradient \\
     {$\triangleright$ sample within fraction to find ${\bar{x}}_{t_e(f)}$} \\  
    \For{$t_o$ in range($t_s(f_o)$, $t_e(f_o)-1$, $-k$)}{
      $\hat{\bar{x}}_{0} \leftarrow \mathbf{D}_{\theta_P}^{[f_o]}(\bar{x}_{t_o})$ \\
      \uIf{$t_o > t_e(f_o)$}{
      $\bar{x}_{t_o-k} \leftarrow \alpha_{t_o-k} \hat{\bar{x}}_{0}   + \sigma_{t_o-k} \frac{\bar{x}_{t_o} - \alpha_{t_o} \hat{\bar{x}}_{0} }{\sigma_{t_o}}$
      }\Else{
     $\gamma \leftarrow s \frac{\sigma_{t_o}^2}{\alpha_{t_o}^2} (\alpha_{t_0 - k} - \alpha_{t_o} \frac{\sigma_{t_o-k}}{\sigma_{t_o}}) $ $\triangleright$ scale \\
             $\bar{x}_{t_o-k} \leftarrow \alpha_{t_o-k} \hat{\bar{x}}_{0} + \sigma_{t_o-k} \frac{\bar{x}_{t_o} - \alpha_{t_o} \hat{\bar{x}}_{0}}{\sigma_{t_o}} + \gamma G$      
      }
    }
}
\end{algorithm}

\section{Methods}
\subsection{Experimental Procedures}  
Demonstrations were performed on fMRI scans from HCP-Rest, HCP-Task \cite{van2013wu} and ID1000 datasets \cite{snoek2021amsterdam}. After exclusion of incomplete scans ($<$1200s), HCP-Rest contained 1093 resting-state fMRI samples (594 female, 499 male), and HCP-Task contained 7450 task-based fMRI scans (594 female, 501 male) where each subject performed 7 different tasks (i.e., emotion, relational, gambling, language, social, motor, working memory). ID1000 contained 881 movie-watching fMRI samples (458 female, 423 male). Brain regions were defined via an anatomical atlas in the MNI template \cite{schaefer2018local}. 

Experiments were conducted on an NVIDIA RTX 3090 GPU via PyTorch. Each dataset was split into training (80\%), validation (10\%) and test sets (10\%) with no subject overlap. Identical data splits were used for all models. A transformer-based downstream fMRI classifier was trained using cross-entropy loss \cite{bedel2023bolt}. For explanation methods, priors were trained using their originally proposed losses, and hyperparameters including number of epochs, learning rate and batch size were selected to minimize validation loss \cite{li2021braingnn}. For each method, a common set of hyperparameters yielding near-optimal performance was used across datasets \cite{elmas2022federated}.

\begin{figure*}[t]
  \begin{minipage}{0.225\textwidth}
    \caption{Explanation maps produced by competing methods for a representative fMRI scan from the HCP-Task dataset. Explanation maps are shown along with an original input sample for the motor task (left) and the subject-average global responses for the motor task (right). For counterfactual methods, counterfactual samples were generated separately to flip the class label from the motor onto each of six remaining cognitive tasks, and the results were averaged. In explanation maps, bluer tones indicate features with higher importance, white tones indicate features with lower importance for the motor task.}
    \label{fig:Attribution}
  \end{minipage}
  \begin{minipage}{0.775\textwidth}
    \centering
    \includegraphics[width=0.95\linewidth]{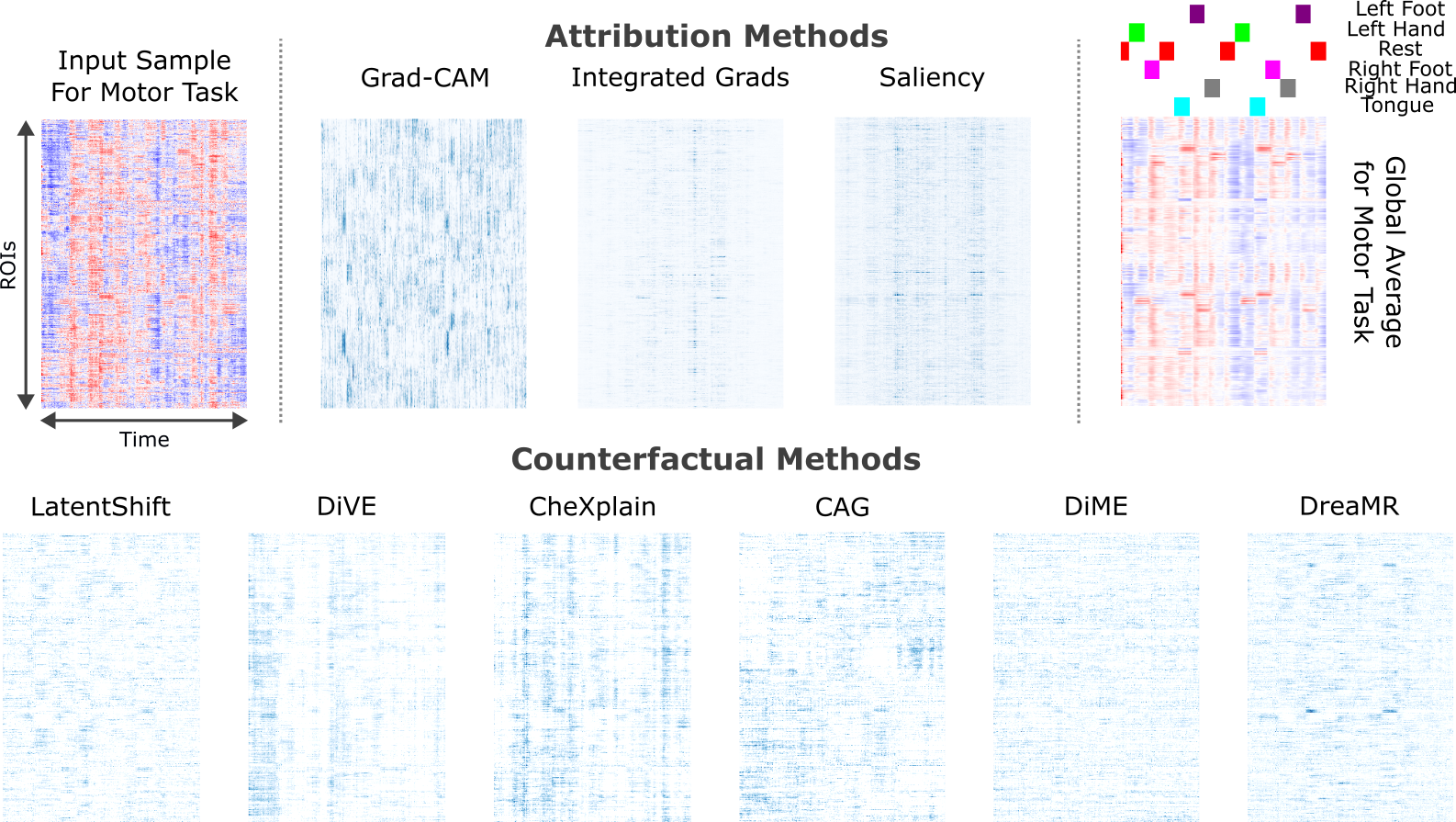}  
  \end{minipage}
    \hfill
\end{figure*}

\begin{table*}[t]
\caption{Performance in counterfactual explanation for HCP-Rest, HCP-Task and ID1000. Proximity, sparsity, FID are listed as mean$\pm$std across the test set. The top-performing method in each dataset is marked in boldface.}
\centering
\resizebox{0.8\textwidth}{!}{%
\begin{NiceTabular}{l|ccc|ccc|ccc}
\hline
\multirow{2}{*}{}          & \multicolumn{3}{c}{HCP-Rest} & \multicolumn{3}{c}{HCP-Task} & \multicolumn{3}{c}{ID1000}  \\ \cline{2-10}
                        & Prox. \textdownarrow      &  Spar. \textdownarrow           & FID \textdownarrow    
                        & Prox. \textdownarrow      &  Spar. \textdownarrow           & FID \textdownarrow    
                        & Prox. \textdownarrow      &  Spar. \textdownarrow           & FID \textdownarrow    
                        \\ \hline
DreaMR & \textbf{38.4$\pm$1.4} & \textbf{10.4$\pm$0.6} & \textbf{22.4$\pm$0.7} & \textbf{53.2$\pm$0.6} & \textbf{16.3$\pm$0.2} & \textbf{12.5$\pm$0.8} & \textbf{44.6$\pm$2.1} & \textbf{11.8$\pm$0.6} & \textbf{25.7$\pm$1.2}  \\ \hline
DiME & 65.2$\pm$1.3 & 20.2$\pm$0.5 & 37.5$\pm$1.0 & 60.1$\pm$0.3 & 18.6$\pm$0.1 & 65.1$\pm$1.2 & 67.2$\pm$0.7 & 21.3$\pm$0.3 & 36.0$\pm$2.6  \\ \hline
CAG & 120.6$\pm$2.2 & 34.1$\pm$0.6 & 27.3$\pm$1.8 & 137.2$\pm$4.0 & 37.5$\pm$0.8 & 18.7$\pm$2.8 & 195.7$\pm$6.8 & 44.6$\pm$1.1 & 30.6$\pm$0.8 \\ \hline
CheXplain & 128.2$\pm$4.4 & 37.7$\pm$0.8 & 114.6$\pm$10.8 & 174.7$\pm$26.6 & 43.3$\pm$3.2 & 94.2$\pm$17.7 & 131.3$\pm$13.4 & 37.3$\pm$2.7 & 92.2$\pm$3.6 \\ \hline
DiVE & 112.6$\pm$1.6 & 33.1$\pm$0.5 & 97.9$\pm$0.5 & 101.7$\pm$0.7 & 30.7$\pm$0.2 & 107.7$\pm$1.0 & 97.3$\pm$0.2 & 29.4$\pm$0.1 & 101.0$\pm$1.4\\ \hline
LatentShift & 45.3$\pm$1.1 & 13.0$\pm$0.5 & 34.9$\pm$1.2 & 92.9$\pm$14.8 & 25.7$\pm$2.0 & 137.9$\pm$2.9 & 115.0$\pm$9.5 & 16.5$\pm$1.1 & 89.1$\pm$8.2   \\ \hline
\end{NiceTabular}
}
\label{tab:allmetrics}
\end{table*}

\subsection{Competing Methods}
DreaMR was compared against state-of-the-art counterfactual methods based on VAE, GAN, and diffusion priors.

\subsubsection{DreaMR}
DreaMR used an efficient transformer architecture for reverse diffusion \cite{bedel2023bolt}. The FMD prior used $T$=1024 steps, $F$=4 fractions, $P$=7 distillation phases, $k$=128 final step size. Cross-validated hyperparameters were E=20 epochs, $\eta$=2x10$^{-4}$ learning rate, B=8 batch size, $s$=(25,50,100) for ID1000, HCP-Task, HCP-Rest respectively. 

\subsubsection{DiME}
A diffusion-based counterfactual method was trained to generate fMRI samples, and counterfactual generation was achieved by interleaved sampling guided by downstream classification loss and perceptual loss between original and resampled fMRI scans \cite{jeanneret2022diffusion}. Cross-validated hyperparameters were E=200 epochs, $\eta$=5x10$^{-5}$, B=8.

\subsubsection{CAG}
A GAN-based counterfactual method was trained to translate between separate classes using adversarial loss and cross-entropy loss from the downstream classifier, counterfactual generation was achieved via translation \cite{matsui2022counterfactual}. Cross-validated hyperparameters were E=200, $\eta$=5x10$^{-5}$, B=32.

\subsubsection{CheXplain}
A GAN-based counterfactual method was trained to generate samples conditioned on predictions by the downstream classifier; counterfactual generation was achieved by modifying style latents in the generator \cite{atad2022chexplaining}. Cross-validated hyperparameters were E=200, $\eta$=2x10$^{-4}$, B=16.

\subsubsection{DiVE}
A VAE-based counterfactual method was trained via a variational objective to generate fMRI samples, and counterfactual generation was achieved by modifying encoded latents of the autoencoder \cite{rodriguez2021beyond}. Cross-validated hyperparameters were E=500, $\eta$=4x10$^{-4}$, B=16.

\subsubsection{LatentShift}
An autoencoder-based counterfactual method that was designed to modify the latent representation of an input image to emphasize or de-emphasize features that contribute to the prediction \cite{cohen2021gifsplanation}. Cross-validated hyperparameters were E=100, $\eta$=4x10$^{-4}$, B=16.

\subsection{Performance Evaluation}
To assess sample fidelity, we quantified proximity, sparsity and Fréchet Inception Distance (FID) metrics between original and counterfactual samples \cite{augustin2022diffusion}. Proximity was taken as the normalized $\ell_2$ distance between samples, and sparsity was taken as the number of features altered between samples. Low proximity and sparsity indicate that feature changes between samples are minimal, so the resultant interpretations are specific. Low FID scores suggest that counterfactual samples are drawn from a similar distribution to that of original samples, so the interpretations are plausible. Significance of differences in performance metrics between competing methods were assessed via non-parametric Wilcoxon signed-rank tests.  %These metrics were also used to compare functional connectivity (FC) matrices derived from original versus counterfactual samples, albeit Earth Mover's Distance (EMD) was adopted for FC instead of FID [WHY?]. 

\begin{table*}[t]
\caption{Performance on functional connectivity (FC) features derived from counterfactual samples for HCP-Rest, HCP-Task and ID1000.}
\centering
\resizebox{0.8\textwidth}{!}{%
\begin{NiceTabular}{l|ccc|ccc|ccc}
\hline
\multirow{2}{*}{}          & \multicolumn{3}{c}{HCP-Rest} & \multicolumn{3}{c}{HCP-Task} & \multicolumn{3}{c}{ID1000}  \\ \cline{2-10}
                        & Prox. \textdownarrow      &  Spar. \textdownarrow           & FID \textdownarrow    
                        & Prox. \textdownarrow      &  Spar. \textdownarrow           & FID \textdownarrow    
                        & Prox. \textdownarrow      &  Spar. \textdownarrow           & FID \textdownarrow    
                        \\ \hline
DreaMR & \textbf{1.3$\pm$0.0} & \textbf{1.2$\pm$0.2} & \textbf{45.2$\pm$3.7} & \textbf{1.9$\pm$0.0} & \textbf{9.6$\pm$0.5} & \textbf{20.1$\pm$0.6} & \textbf{2.4$\pm$0.2} & \textbf{8.2$\pm$0.9} & \textbf{56.2$\pm$2.3}  \\ \hline
DiME & 6.7$\pm$0.2 & 19.1$\pm$0.2 & 96.6$\pm$5.1 & 9.0$\pm$0.2 & 48.6$\pm$1.2 & 85.1$\pm$1.8 & 3.7$\pm$0.1 & 13.7$\pm$0.8 & 86.7$\pm$8.3\\ \hline
CAG & 16.6$\pm$0.8 & 39.9$\pm$0.5 & 69.6$\pm$1.3 & 7.9$\pm$0.1 & 36.9$\pm$0.6 & 23.4$\pm$1.7 & 12.5$\pm$0.1 & 41.5$\pm$0.4 & 57.4$\pm$1.0 \\ \hline
CheXplain & 17.4$\pm$2.0 & 43.2$\pm$3.4 & 173.9$\pm$7.6 & 10.1$\pm$3.2 & 41.6$\pm$3.0 & 172.2$\pm$54.5 & 14.6$\pm$0.4 & 49.2$\pm$1.8 & 186.0$\pm$32.6 \\ \hline
DiVE & 34.3$\pm$2.8 & 66.8$\pm$2.0 & 154.7$\pm$6.1 & 23.4$\pm$1.3 & 75.2$\pm$1.7 & 146.8$\pm$2.0 & 24.9$\pm$1.3 & 69.6$\pm$2.4 & 164.6$\pm$2.7\\ \hline
LatentShift & 9.3$\pm$0.6 & 26.9$\pm$0.7 & 72.0$\pm$1.2 & 9.6$\pm$1.0 & 42.8$\pm$2.2  & 76.7$\pm$4.3 & 8.6$\pm$0.1 & 33.6$\pm$1.1 & 86.4$\pm$7.01   \\ \hline
\end{NiceTabular}
}
\label{tab:fcmetrics}
\end{table*}

\begin{table}[t]
    \caption{Performance of a logistic classifier trained using counterfactual samples from competing methods. Accuracy (\%) and F1 (\%) metrics are listed across the test set.}
\centering
\resizebox{0.81\columnwidth}{!}{
\begin{NiceTabular}{l|cc|cc|cc}
    \hline
    \multirow{2}{*}{} & \multicolumn{2}{c}{HCP-Rest} & \multicolumn{2}{c}{HCP-Task} & \multicolumn{2}{c}{ID1000} \\
    \cline{2-7}& Acc. \textuparrow & F1 \textuparrow & Acc. \textuparrow & F1 \textuparrow & Acc. \textuparrow & F1 \textuparrow \\
    \hline
    DreaMR & \textbf{74.0} & \textbf{75.5} & \textbf{85.8} & \textbf{86.2} & \textbf{68.7} & \textbf{69.3} \\ \hline
    DiME & 67.6 & 73.8 & 83.5 & 83.6 & 65.9 & 68.0 \\ \hline
    CAG & 49.1 & 64.1 & 58.3 & 58.6 & 55.1 & 67.9 \\ \hline
    CheXplain & 66.4 & 71.8 & 6.7 & 13.3 & 55.7 & 58.3 \\ \hline
    DiVE & 73.6 & 71.8 & 66.1 & 64.9 & 67.3 & 66.8 \\ \hline
    LatentShift & 69.1 & 67.9 & 76.1 & 76.3 & 61.4 & 63.4 \\ \hline    
\end{NiceTabular}
}
\label{tab:biomarkerPerf}
\end{table}

\section{Results}

\subsection{Counterfactual Generation}
\label{sec:generation}
We first demonstrated DreaMR in explanation of downstream transformer-based classifiers for gender on HCP-Rest and ID1000, and for cognitive task on HCP-Task. Fig. \ref{fig:Attribution} depicts representative counterfactual samples from competing methods along with heatmaps from attribution methods \cite{lin2022sspnet,sundararajan2017axiomatic}. Among attribution methods, Grad-CAM suffers from poor specificity, whereas integrated gradients and saliency methods show low sensitivity to global context. Although counterfactual methods can maintain relatively high specificity and contextual sensitivity, sample fidelity is limited for methods based on VAE, GAN priors. In contrast, DreaMR generates a high-fidelity counterfactual that is more closely aligned with prominent features in the average samples recorded for the target cognitive task, even against the diffusion-based DiME method. These results suggest that DreaMR offers higher specificity and fidelity than competing methods in counterfactual generation.

Next, counterfactual samples were quantitatively evaluated for their specificity as measured by proximity and sparsity, and for their fidelity as measured by FID. Table \ref{tab:allmetrics} lists performance metrics averaged across classes on each dataset. Overall, we find that DreaMR achieves the top performance in all datasets (p$<$0.05). On average, DreaMR outperforms competing methods in proximity by 56.0, sparsity by 17.2, FID by 40.0 on HCP-Rest; proximity by 60.1, sparsity by 14.9, FID by 72.2 on HCP-Task; and proximity by 76.7, sparsity by 18.0, FID by 44.1 on ID1000. These results indicate that DreaMR generates plausible counterfactual samples that are closely aligned with the data distribution, and that have relatively small differences with the original samples.

A pervasively analyzed attribute in fMRI studies is functional connectivity (FC) that is taken as response correlation between brain regions, and FC features are commonly associated with cognitive states \cite{pereira2009machine}. Thus, we also examined the specificity and fidelity of FC features derived from counterfactual fMRI samples. Table \ref{tab:fcmetrics} lists performance metrics based on FC features on each dataset. Overall, we find that DreaMR achieves the top performance in all datasets (p$<$0.05). On average, DreaMR outperforms competing methods in proximity by 15.6, sparsity by 38.0, FID by 68.2 on HCP-Rest; proximity by 10.1, sparsity by 39.4, FID by 80.7 on HCP-Task; and proximity by 10.5, sparsity by 33.3, FID by 60.0 on ID1000. These results indicate that DreaMR generates plausible counterfactual samples whose FC features are closely aligned to those of original fMRI samples.

\begin{figure*}[t]
    \begin{minipage}{0.34\textwidth}
        \caption{Biomarker maps produced by DreaMR for female (top panel) and male (bottom panel) genders on the HCP-Rest dataset. To obtain biomarker maps, counterfactual samples were generated for each original fMRI sample to flip the decisions of a deep fMRI classifier for gender detection. FC features of original and counterfactual samples were separately derived, and the differences between the two sets of samples were averaged across subjects. Important FC features were determined by selecting the features showing the top 5\% of differences. Each important ROI is marked with a dot on the anatomical template, and connectivity between ROIs is shown with a bar. Dot size denotes the importance of the ROI, and bar thickness denotes the importance of the connection. ROIs are colored according to the functional network they belong to (see legend). Lateral and superior brain views are displayed. LH: left hemisphere, RH: right hemisphere; SomMot: somatomotor; DorsAttn: dorsal attention; SalventAttn: Salience/ventral attention.}
        \label{fig:biomarker_hcprest}
    \end{minipage}
    \begin{minipage}{0.66\textwidth}
        \centering
        \includegraphics[width=0.95\columnwidth]{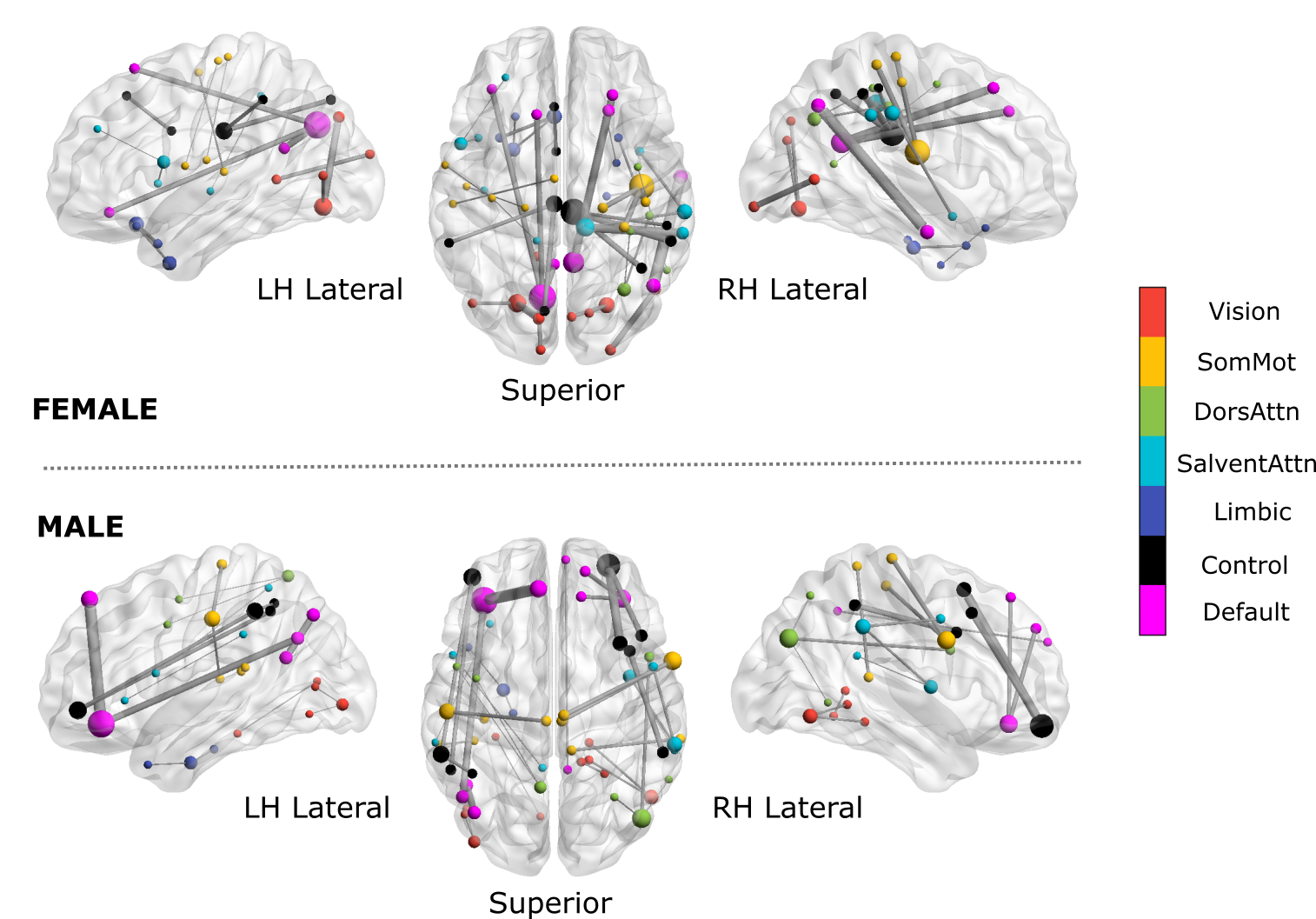}  
    \end{minipage}
    \hfill
\end{figure*}
\subsection{Biomarker Identification}
Neuroscience literature suggests that individual cognitive states are associated with characteristic response patterns across brain regions \cite{pereira2009machine,ccukur2013attention}. In principle, the difference between a counterfactual and an original fMRI sample reflects the difference between characteristic patterns for target versus original cognitive states. Thus, when a classifier is used to discriminate between the two states given the differences between counterfactual and original samples, classification performance should grow as the quality of counterfactuals improves. To examine this issue, we built a separate logistic classifier using the counterfactual samples generated by each explanation method \cite{bedel2023bolt}. The logistic classifiers were tested on the original fMRI samples to perform the same tasks as deep classifiers. Table \ref{tab:biomarkerPerf} lists accuracy and F1 for competing methods. DreaMR yields the highest performance in all cases (p$<$0.05). On average, DreaMR outperforms competing methods in accuracy by 8.8\%, F1 by 5.6\% on HCP-Rest; accuracy by 27.7\%, F1 by 26.9\% on HCP-Task; accuracy by 7.6\%, F1 by 4.4\% on ID1000. This finding indicates that DreaMR generates counterfactuals that better capture the characteristic differences in response patterns for distinct cognitive states. 

Next, counterfactual samples generated by DreaMR were interpreted to derive biomarker maps that reflect the important FC features associated with cognitive state. To do this, the differences between the FC features of original and counterfactual samples were derived. Important FC features showing the top 5\% of differences on average across subjects were determined. Representative biomarker maps for each gender on HCP-Rest are displayed in Fig. \ref{fig:biomarker_hcprest}. For females, important FC features are observed in posterior segments of default mode and control networks, vision network, and right lateralized somatomotor network. Instead, for males, important FC features are observed in frontal segments of the default and control networks, dorsal attention network, and bilateral somatomotor network. These findings are closely aligned with previously reported resting-state connectivity patterns for females and males \cite{ritchie2018sex,kim2021learning}. Thus, DreaMR offers a promising framework to identify imaging biomarkers characteristic to individual cognitive states.  

%MAles: frontal segments of the default network, 
%No vision
%bilateral somatomotor
%frontal control network
%dorsal attention network

%Females: posterior segments of the default network
%right laterialized somatomoro
%posterior control network
%vision network
%no dorsal attention

\subsection{Computational Efficiency}
A practical concern for counterfactual methods is their efficiency in resampling of original fMRI samples while trying to flip the classifier decision. Table \ref{tab:efficiency} lists the inference time and memory use of competing methods. In terms of inference time, the non-iterative CAG method is the fastest, whereas the iterative DiME method based on a conventional diffusion prior is the slowest. The remaining iterative methods including DreaMR require intermediate inference times between the two extremes. Note that the FMD prior in DreaMR enables substantially improved efficiency over the conventional diffusion prior, and comparable efficiency with VAE and GAN priors. In terms of memory load, DreaMR has comparable demand to methods based on VAE and GAN priors, and it has significantly lower memory demand than DiME that uses a relatively complex UNet architecture. 

\begin{table}[t]
    \caption{Inference times (msec) and memory load (gigabytes) per generation of a counterfactual fMRI sample.}
        \renewcommand*{\arraystretch}{1.2}
\centering
\resizebox{0.80\columnwidth}{!}{
\begin{NiceTabular}{l|cc|cc|cc}
    \hline
    \multirow{2}{*}{} & \multicolumn{2}{c}{HCP-Rest} & \multicolumn{2}{c}{HCP-Task} & \multicolumn{2}{c}{ID1000} \\
    \cline{2-7}& Inf.  & Mem.  & Inf.  & Mem. & Inf.  & Mem.  \\
    \hline
    DreaMR & 2735 & 3.7 & 1173 & 2.7 & 1114 & 2.7 \\ \hline
    DiME & 22104 & 6.0 & 14806 & 4.2 & 14918 & 4.2 \\ \hline
    CAG & 116 & 4.3 & 45 & 3.2 & 45 & 3.3 \\ \hline
    CheXplain & 384 & 3.0 & 491 & 2.9 & 282 & 2.9 \\ \hline
    DiVE & 1750 & 3.8 & 1121 & 3.1 & 1149 & 2.8 \\ \hline
    LatentShift & 967 & 3.7 & 850 & 2.9 & 829 & 2.2 \\ \hline    
\end{NiceTabular}
}
\label{tab:efficiency}
\end{table}

% \begin{table}[t]
%     \caption{Inference times (sec) and memory load (gigabytes) per generation of a counterfactual fMRI sample.}
%         \renewcommand*{\arraystretch}{1.2}
% \centering
% \resizebox{0.75\columnwidth}{!}{
% \begin{NiceTabular}{l|cc|cc|cc}
%     \hline
%     \multirow{2}{*}{} & \multicolumn{2}{c}{HCP-Rest} & \multicolumn{2}{c}{HCP-Task} & \multicolumn{2}{c}{ID1000} \\
%     \cline{2-7}& Inf.  & Mem.  & Inf.  & Mem. & Inf.  & Mem.  \\
%     \hline
%     DreaMR & 3.2 & 3.7 & 1.1  & 2.7 & 1.4 & 2.7 \\ \hline
%     DiME & 23.4 & 6.0 & 15.3 & 4.2 & 16.0 & 4.2 \\ \hline
%     CAG & 0.9 & 4.3 & 0.1 & 3.2 & 0.8 & 3.3 \\ \hline
%     CheXplain & 0.4 & 3.0 & 32.6 & 2.9 & - & 2.9 \\ \hline
%     DiVE & 1.9 & 3.8 & 1.3 & 3.1 & 1.3 & 2.8 \\ \hline
%     LatentShift & 1.1 & 3.7 & 0.1 & 2.9 & 0.9 & 2.2 \\ \hline    
% \end{NiceTabular}
% }
% \label{tab:efficiency}
% \end{table}

\subsection{Ablation Studies}
A series of ablation studies were conducted on the HCP-Rest dataset to assess the importance of main design elements in DreaMR. First, we trained several ablated variants of DreaMR to assess the influence of the transformer architecture, the denoised sample estimate for calculating classifier guidance, and multi-phase distillation. A `w/o transformer' variant was formed by replacing the transformer with the common UNet architecture for diffusion priors. A `w/o Langevin' variant was formed by replacing the denoised sample estimate obtained through Langevin sampling across diffusion fractions with a single-shot estimate obtained via the denoising network of the current fraction. A `w/o multi-phase' variant was formed by performing a single-phase distillation of the diffusion prior. %A `w/o incremental' variant was formed by adopting DDPM sampling in the diffusion prior. 
Table \ref{tab:ablation_others} lists performance for DreaMR and ablated variants. Overall, we find that DreaMR outperforms all ablated variants, indicating that each interrogated design element contributes to method performance.

Next, we examined the influence of the number of diffusion fractions ($F$) and distillation phases ($P$) on the counterfactual samples generated by the FMD prior. Table \ref{tab:ablation_fmd} lists performance metrics for FMD priors obtained for varying $F$ while $P$=7, and for varying $P$ while $F$=4. We find that the selected values of $F$=4, $P$=7 generally attain near-optimal performance. Although $F$=8 yields slightly lower proximity and sparsity values, $F$=4 is preferable as it yields lower FID indicating higher sample fidelity and it requires fewer denoising networks.

\section{Discussion}
DreaMR is a novel explanation method for fMRI based on an efficient FMD prior that generates high-quality counterfactuals in brief sampling times. Experimental demonstrations indicate the superior performance of DreaMR against competing counterfactual methods based on VAE, GAN and conventional diffusion priors. In particular, the difference between original and counterfactual samples produced by DreaMR enables more specific and plausible explanations than baselines.

Here, we utilized DreaMR to explain classifiers that predict discrete cognitive states given brain responses via a transformer model. Several extensions can be pursued to expand the scope of the proposed methodology. First, classifiers based on alternative convolutional or recurrent architectures might be considered \cite{kawahara2017brainnetcnn,li2021braingnn,kim2021learning}. Since counterfactual generation is a model-agnostic framework, DreaMR can in principle be adopted to other architectures without modification. Second, many neurodegenerative and neurodevelopmental diseases are reported to have complementary biomarkers in fMRI as well as anatomical or diffusion-weighted MRI \cite{fan2007multivariate,zhang2021fusion}. A more performant classifier for disease-related cognitive states could be attained by using multi-modal images as input. DreaMR can be adopted for explaining such multi-modal classifiers by training a multi-modal FMD prior.  

It might also be possible to employ DreaMR for explaining regression models. The human brain represents information on continuous stimulus or task variables \cite{ccukur2013attention}, which can be decoded from brain responses via regression \cite{vanrullen2019reconstructing}. Deep regression models can also be employed to predict voxel- or ROI-wise responses given stimulus variables \cite{ngo2022transformer}. To adopt DreaMR, guidance from classification loss during counterfactual generation could be replaced with guidance from regression losses. It remains important future work to assess the efficacy of DreaMR in a broader set of explanation tasks.

\begin{table}[t]
\caption{Performance metrics for ablated variants of DreaMR without a transformer, without Langevin sampling, and without multi-phase distillation.}
\centering
\resizebox{0.7\columnwidth}{!}{%
\begin{NiceTabular}{lccc}
\hline
 & Prox. \textdownarrow      &  Spar. \textdownarrow           & FID \textdownarrow   \\ \hline
DreaMR & \textbf{38.7$\pm$1.9} & \textbf{10.6$\pm$0.8} & \textbf{21.5$\pm$0.4} \\ \hline
w/o transformer & 47.5$\pm$2.3 & 14.0$\pm$0.9 & 33.1$\pm$1.0 \\ \hline
w/o Langevin & 44.7$\pm$2.1 & 13.0$\pm$0.8 & 22.5$\pm$0.7 \\ \hline
w/o multi-phase & 40.7$\pm$0.9 & 11.4$\pm$0.4 & 22.6$\pm$0.1 \\ \hline
%w/o incremental & a$\pm$b & a$\pm$b & a$\pm$b \\ \hline
\end{NiceTabular}
}
\label{tab:ablation_others}
\end{table}

\begin{table}[t]
\caption{Performance for the FMD prior implemented with varying number of fractions ($F$) and number of distillation phases ($P$).}
\centering
\setlength{\tabcolsep}{2pt} % Horizontal Padding
\resizebox{0.95\columnwidth}{!}{%
\begin{NiceTabular}{cccc|cccc}
\hline
 \multicolumn{4}{c}{Fractions} & \multicolumn{4}{c}{Distillation phases}  \\ \hline $F$ & Prox.\textdownarrow     &  Spar. \textdownarrow           & FID \textdownarrow    & $P$ & Prox.\textdownarrow      & Spar. \textdownarrow  & FID \textdownarrow \\ \hline
1 & 42.0$\pm$2.0 & 12.0$\pm$0.9 & 22.0$\pm$0.1 & 5 & 41.5$\pm$1.6 & 11.7$\pm$0.7 & 20.8$\pm$1.2 \\ \hline
2 & 40.3$\pm$1.8 & 11.2$\pm$0.7 & 21.9$\pm$0.7 & 6 & 40.1$\pm$1.6 & 11.2$\pm$0.7 & 21.3$\pm$0.3 \\ \hline
4 & 38.7$\pm$1.9 & 10.6$\pm$0.8 & \textbf{21.5$\pm$0.4} & 7 & \textbf{38.7$\pm$1.9} & \textbf{10.6$\pm$0.8} & \textbf{21.5$\pm$0.4} \\ \hline
8 & \textbf{38.0$\pm$2.0} & \textbf{10.2$\pm$0.8} & 22.1$\pm$1.3 & 8 & 45.1$\pm$2.3 & 12.8$\pm$0.8 & 23.7$\pm$0.5 \\ \hline
\end{NiceTabular}
}
\label{tab:ablation_fmd}
\end{table}

\begin{comment}
\begin{table}[t]
\caption{Proximity, sparsity, WD of functional connectivity matrices derived from counterfactual fMRI samples.}
\centering
\resizebox{0.67\columnwidth}{!}{%
\begin{NiceTabular}{lccc}
\hline
 & Prox.\textdownarrow      &  Spar.\textdownarrow           & WD \textdownarrow   \\ \hline
DreaMR & \textbf{1.3$\pm$0.0} & \textbf{1.2$\pm$0.2} & \textbf{32.4$\pm$0.3} \\ \hline
DiME & 6.4$\pm$0.2 & 19.1$\pm$0.2 & 72.0$\pm$0.9 \\ \hline
CAG & 16.6$\pm$0.8 & 39.9$\pm$0.5 & 108.6$\pm$2.7 \\ \hline
CheXplain & 17.4$\pm$2.0 & 43.2$\pm$3.4 & 109.9$\pm$6.1 \\ \hline
DiVE & 34.3$\pm$2.8 & 66.8$\pm$2.0 & 162.6$\pm$6.9 \\ \hline
LatentShift & 9.3$\pm$0.6 & 26.9$\pm$0.7 & 85.1$\pm$2.6 \\ \hline
\end{NiceTabular}
}
\label{tab:ablation_FC}
\end{table}
\end{comment}

Several lines of technical limitations could be addressed to further improve DreaMR. Following common practice, here we analyzed fMRI data preprocessed to register brain volumes onto an anatomical template and defined brain regions via an atlas. This procedure facilitates comprehensive and consistent region definitions across subjects \cite{singleton2009functional}, albeit registration onto a template can yield spatial information losses. It might be possible to mitigate potential losses by backprojecting atlas-based region definitions onto the brain spaces of individual subjects \cite{kiremitci2021cocktail}, or by incorporating hybrid spatial encoding based on convolution and self-attention layers to align representations from separate subjects in a common space \cite{malkiel2021pre,dalmaz2022resvit}. 

Second, the diffusion priors in DreaMR were trained here from scratch on public fMRI datasets comprising several hundred subjects. Literature suggests that such sizable datasets might be critical in adequate training of diffusion priors. In resource-limited application domains where data are scarce, fine tuning diffusion priors pretrained on time-series data such as audio waveforms might help improve learning \cite{kong2021diffwave}. Knowledge distillation from adversarial priors might also be employed to facilitate training of diffusion priors \cite{ozbey2022unsupervised}.  

Lastly, DreaMR shortens diffusion sampling times by leveraging fractional multi-phase distillation. Naturally, computational load of counterfactual generation will scale with the duration of fMRI scans. In applications where the load becomes excessive, the FMD prior might be combined with alternative approaches to accelerate diffusion sampling such as consistency \cite{song2023consistency} or adversarial diffusion models \cite{dar2022adaptive}. Further work is warranted to systematically assess the benefits of various acceleration approaches in terms of sampling fidelity versus efficiency during counterfactual generation.

\section{Conclusion}
In this study, we introduced a novel explanation method for fMRI based on diffusion-driven generation of counterfactual samples. Demonstrations on resting-state and task-based fMRI indicate that DreaMR, with its fractional multi-phase distilled diffusion prior, achieves higher sampling efficiency and fidelity against competing methods. Furthermore, it yields enhanced specificity and plausibility in explanations of downstream fMRI classifiers. Therefore, the proposed method holds great promise in enabling sensitive and explainable analysis of multi-variate fMRI data with deep-learning models.

\bibliographystyle{IEEEtran}
\bibliography{IEEEabrv,references}
\end{document}